\documentclass[3p,times,twocolumn]{elsarticle}

\usepackage{ecrc}

\volume{00}

\firstpage{1}

\journalname{Nuclear Physics B Proceedings Supplement}

\runauth{}

\jid{nuphbp}

\jnltitlelogo{Nuclear Physics B Proceedings Supplement}

\usepackage{amssymb}

\usepackage[figuresright]{rotating}

\newcommand{\oh}{\Omega h^2}
\newcommand{\gev}{\mathrm{GeV}}

\newcommand{\sv}{\langle\sigma v\rangle}
\newcommand{\ma}{M_{A^0}}
\newcommand{\mh}{M_{H^0}}
\newcommand{\mhc}{M_{H^\pm}}
\newcommand{\dmn}{M_{N_i}-M_{H^0}}
\newcommand{\mn}[1]{M_{N_{#1}}}

\begin{document}

\begin{frontmatter}



\dochead{}

\title{Implications of the Higgs discovery on minimal dark matter}


\author{M. Klasen}

\address{Institut f\"ur Theoretische Physik, Westf\"alische
 Wilhelms-Universit\"at M\"unster, Wilhelm-Klemm-Stra\ss{}e 9,
 D-48149 M\"unster, Germany}





\begin{abstract}
 The existence of dark matter provides compelling evidence for physics beyond
 the Standard Model. Minimal extensions of the Standard Model with additional
 scalars or fermions allow to explain the observed dark matter relic density in
 an economic way. We analyse several of these possibilities like the inert
 Higgs and radiative seesaw models in the light of the recent Higgs
 discovery and study prospects for the direct and indirect detection of
 dark matter in these models.
\end{abstract}


\end{frontmatter}



\vspace*{-13.5cm}
\noindent MS-TP-14-28
\vspace*{12.05cm}

\section{Motivation}
\label{}

Gravitational effects of dark matter have been observed in galaxies, clusters of galaxies,
the large scale structure of the Universe and the cosmic microwave background radiation. These
observations indicate that dark matter accounts for about $85\%$ of the matter density in the
Universe and for $23\%$ of its total energy density. It must therefore be considered today to
be one of the essential ingredients of our Universe.


Viable dark matter particles should be neutral, stable and weakly interacting, and, to be
consistent with the observed large scale structure of the Universe, behave as \emph{cold}
dark matter. Since none of the Standard Model particles satisfies these conditions, dark
matter provides strong evidence for new physics, and indeed most extensions of the Standard
Model include dark matter candidates. 

WIMP (Weakly Interacting Massive Particle) dark matter represents a generic scenario,
that can naturally account for the observed dark matter density via freeze-out in the early
Universe. Here, the dark matter candidate is a weakly interacting particle with a mass around
the TeV scale - the same scale that is currently being probed by the Large Hadron Collider
(LHC) at CERN.~\footnote{Alternatively, the interactions of the dark matter particles could also
be so weak that they never reach thermal equilibrium, leading to a so-called freeze-in scenario
\cite{Klasen:2013ypa}.}

\section{Minimal models of dark matter}
\label{}

The idea behind minimal models of dark matter is to extend the Standard Model in a minimal
way, so that dark matter can be explained. Typically, these models feature a small number of
additional fields and a new discrete symmetry that stabilises the dark matter particle. They
include models such as the inert doublet model \cite{Klasen:2013btp}, the radiative seesaw model
\cite{Klasen:2013jpa}, and the singlet fermion model \cite{Esch:2013rta}. The coexistence
of two dark matter particles is yet another possibility that is currently being
explored \cite{Esch:2014jpa}, as are coannihilations of dark matter and other
particles \cite{Klasen:2013jpa,Herrmann:2014kma,Harz:2012fz,Herrmann:2011xe}.

\subsection{The inert doublet model}
\label{}

In this model, the Standard Model is extended with an additional scalar doublet, $H_2$, which
is assumed to be odd under a discrete $Z_2$ symmetry. The dark matter candidate is the
neutral component of this new doublet ($H^0$) 
and is then a WIMP featuring gauge and scalar interactions.

\begin{figure}
 \centering
\includegraphics[width=.5\columnwidth]{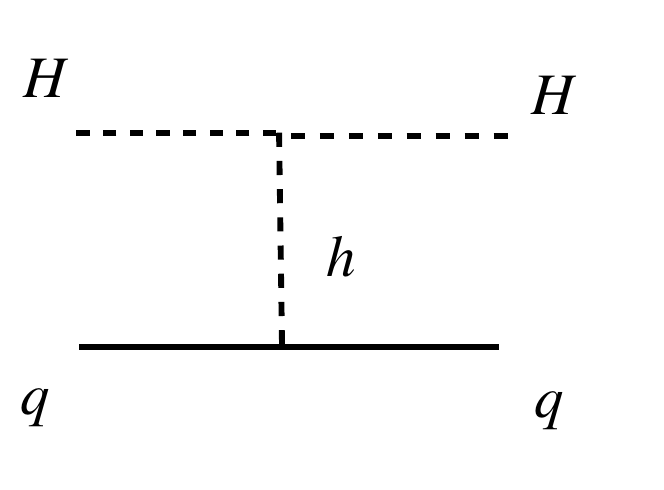}
 \caption{\label{fig:1}Tree-level Feynman diagram for the direct detection cross section of
 inert higgs dark matter.}
\end{figure}

At the tree level (see Fig.\ \ref{fig:1}), the $q$-$H^0$ scattering relevant for direct detection
%
%
proceeds via a Higgs-mediated diagram and is determined by a scalar coupling.
Following the recent LHC discovery of a Standard-Model like Higgs boson, we set $m_{h^0}=125$
GeV.

\begin{figure}
 \centering
\includegraphics[trim= 150mm 0mm 0mm 0mm, clip, width=\columnwidth]{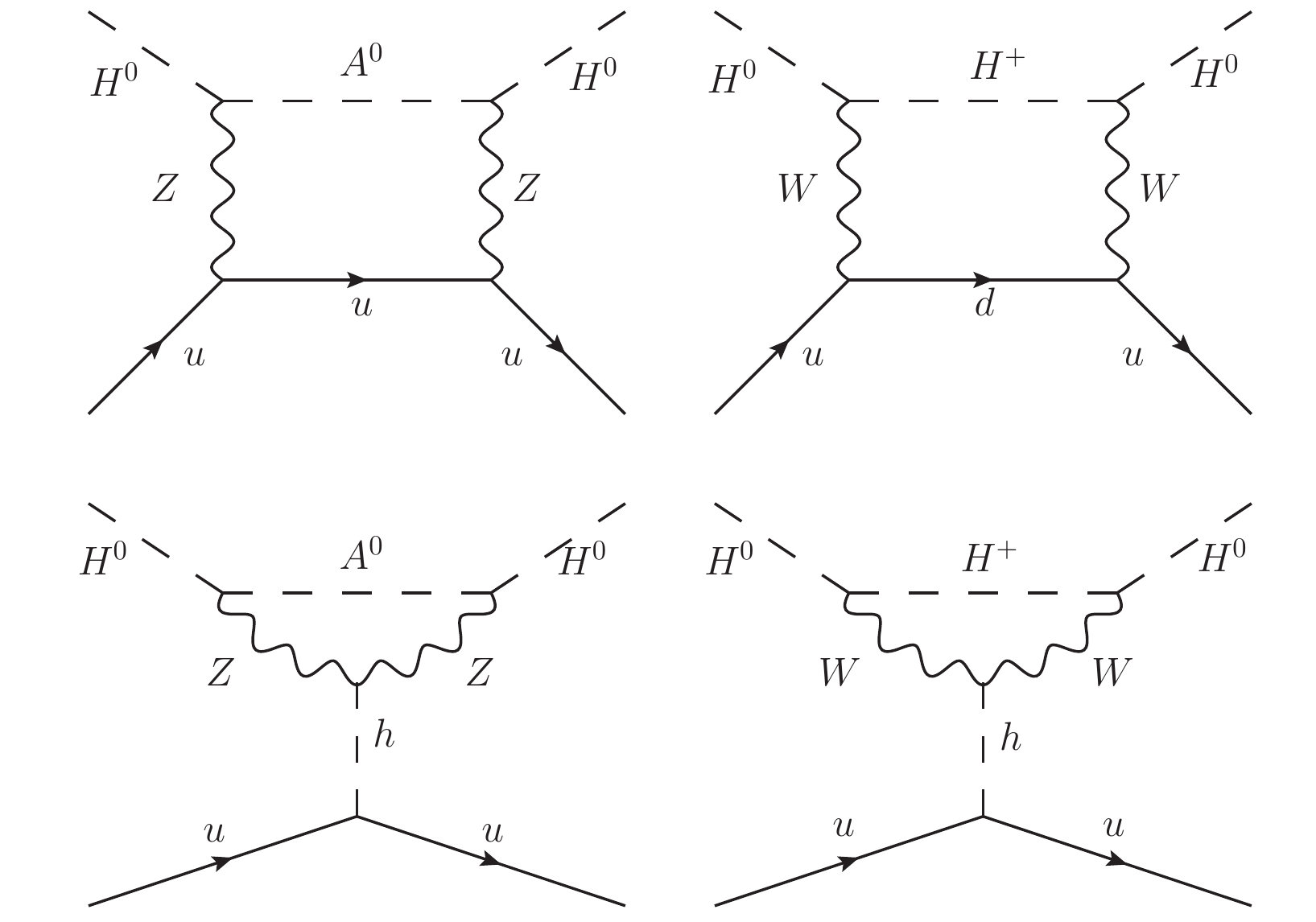}
 \caption{\label{fig:2}One-loop Feynman diagrams that give the dominant corrections to the
 direct detection cross section of inert higgs dark matter \cite{Klasen:2013btp}.}
\end{figure}

\begin{figure}
 \centering
 \includegraphics[width=1.0\columnwidth]{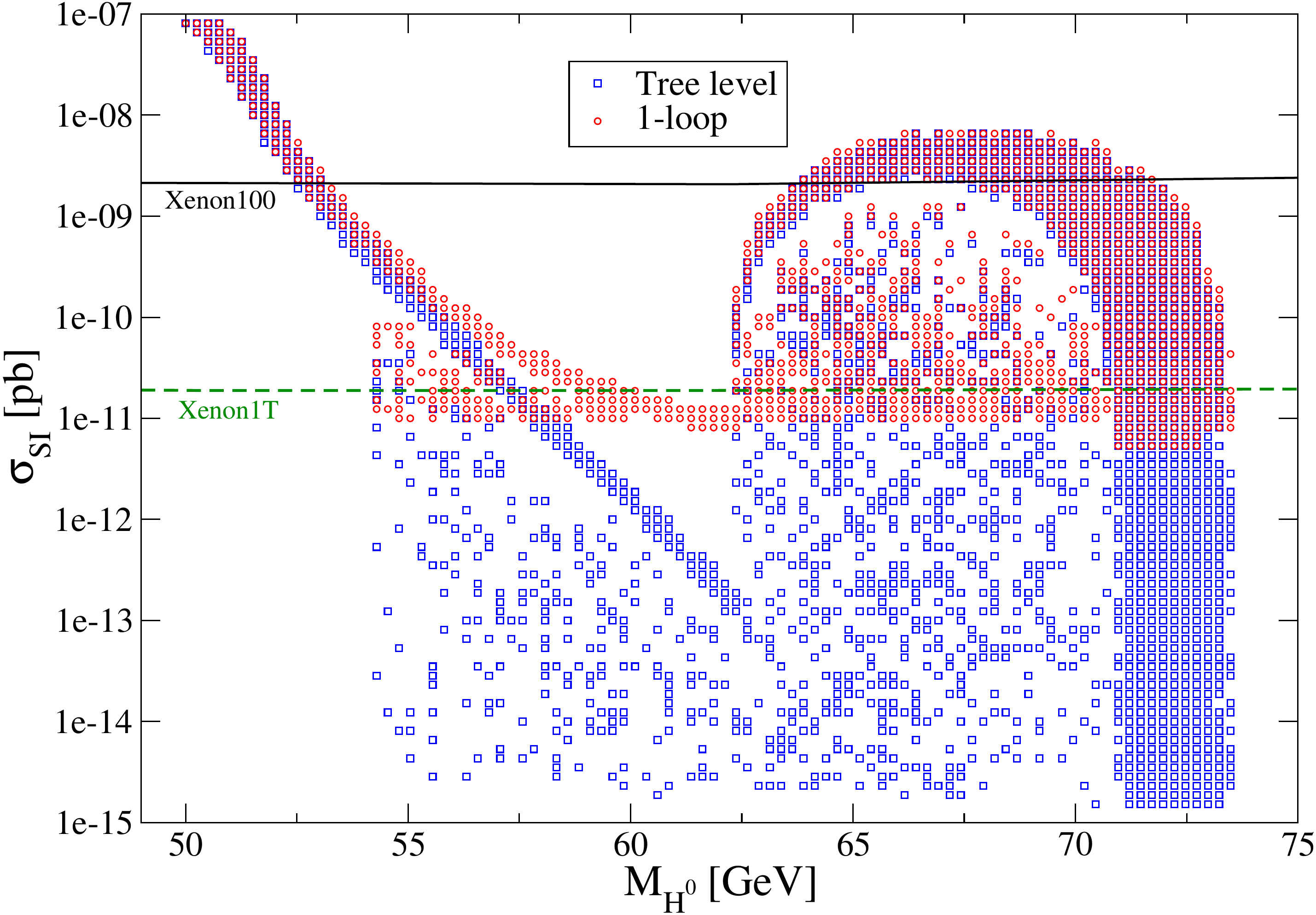}
 \caption{\label{fig:3}Scatter plot of the spin-independent direct detection cross section at
 tree-level and at one-loop as a function of the dark matter mass. In this figure all the
 parameters of the inert higgs model were allowed to vary randomly and all experimental bounds
 were taken into account \cite{Klasen:2013btp}.}
\end{figure}

At the one-loop level (see Fig.\ \ref{fig:2}), the spin-independent direct detection cross
section receives new contributions from $W$- and $Z$-mediated diagrams which are determined by
the gauge couplings.
%
%
%
As a result, the one-loop contribution can actually dominate the direct detection cross section.
In fact, it provides a lower bound on the spin-independent cross section that is within the
reach of planned experiments such as XENON1T \cite{Aprile:2012zx} (see Fig.\ \ref{fig:3})
\cite{Klasen:2013btp}.
%
%

\subsection{The radiative seesaw model}
\label{}

The radiative seesaw model is an extension of the inert doublet model by three singlet fermions
$N_i$ that are odd under the $Z_2$. Its Lagrangian includes the following terms:
\begin{equation}
 {\cal L}=-\frac{M_i}{2}\bar N_i^cP_RN_i+h_{\alpha i}\bar \ell_\alpha H_2^\dagger P_RN_i+{\rm h.c.}
\end{equation}
The main feature of this model is  that it can account also for neutrino masses. They  are
generated at one loop and are given by 
 \begin{equation}
 \left(m_\nu\right)_{\alpha\beta}\simeq \sum_{i=1}^3\frac{2 \lambda_5 h_{\alpha i}h_{\beta i} v^2}{(4\pi)^2 M_i} I\left(\frac{M_i^2}{M_0^2}\right).
\end{equation}     

\begin{figure}
 \centering
 \includegraphics[width=1.0\columnwidth]{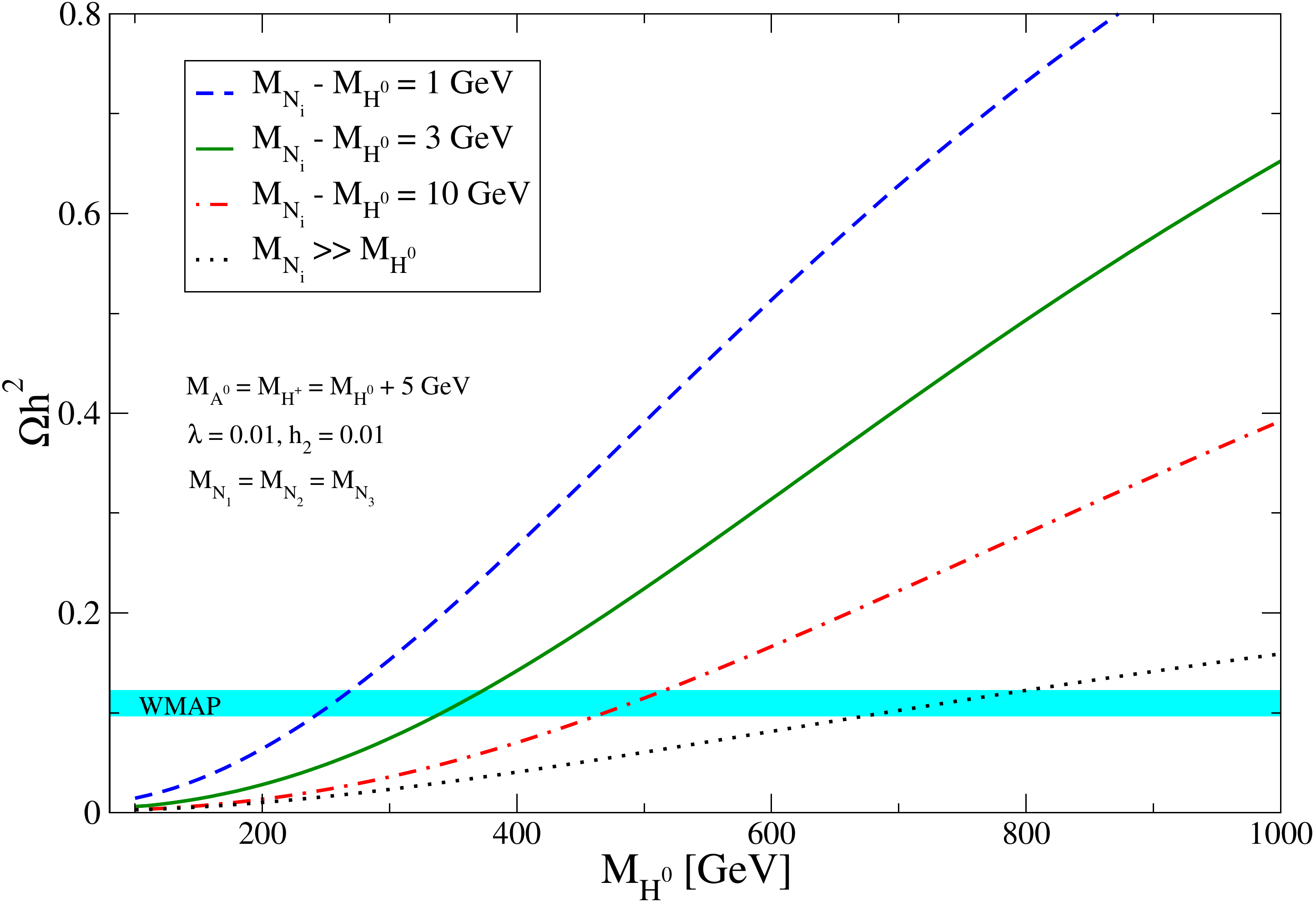}
 \caption{\label{fig:4}The relic density as a function of $\mh$ for different values of $\dmn$.
 In this figure we have set $\lambda=0.01$, $h_2=0.01$, $\ma=\mhc=\mh+5~\gev$ and we have assumed
 that the three fermions have the same mass: $\mn{1}=\mn{2}=\mn{3}$.
 Notice that $\oh$ decreases with increasing $\dmn$ \cite{Klasen:2013jpa}.}
\end{figure}

\begin{figure}
 \centering
 \includegraphics[width=1.0\columnwidth]{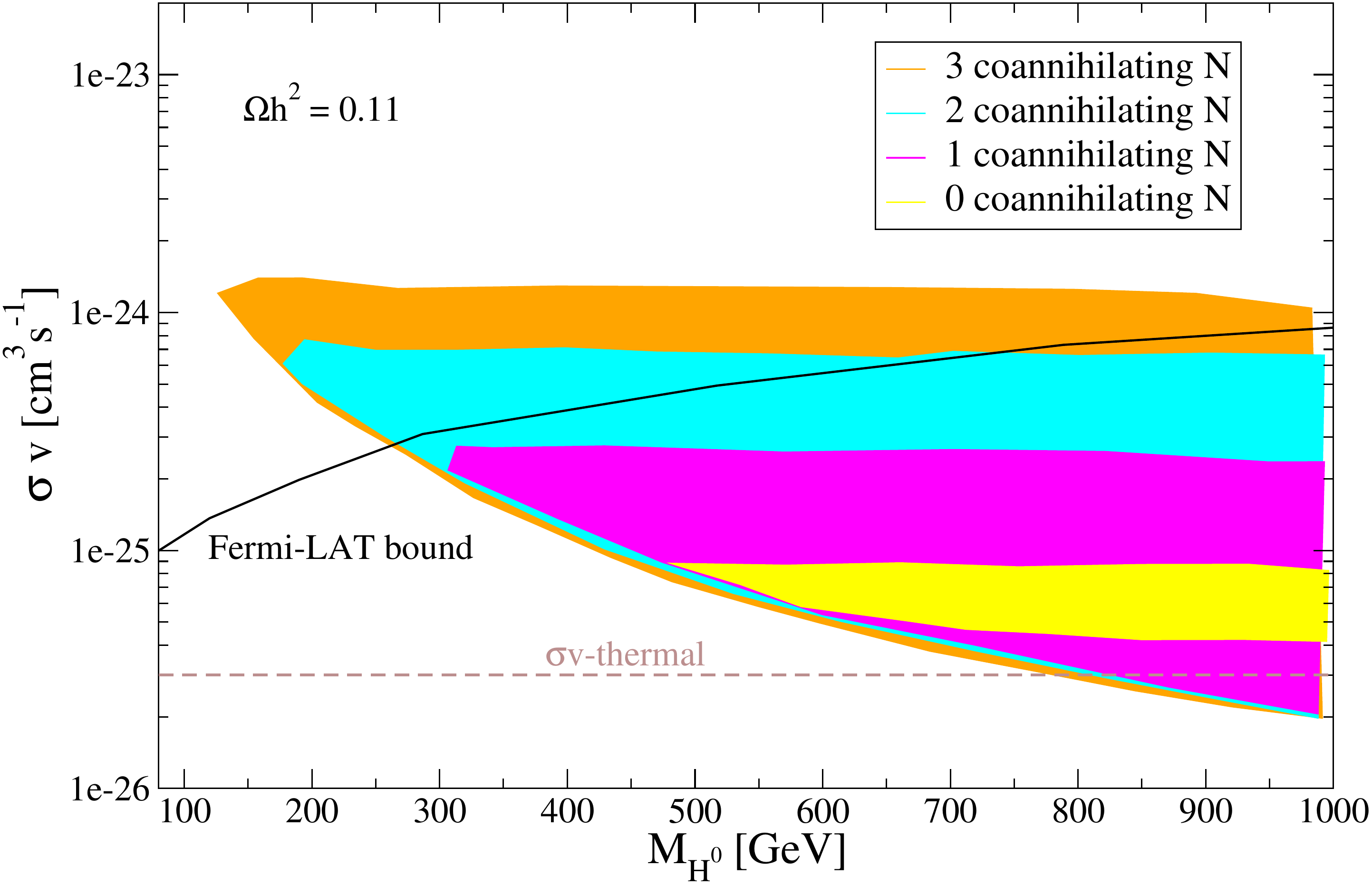}
 \caption{\label{fig:5}Regions in the plane ($\mh$,$\sv$) that are consistent with the dark
 matter constraint for different numbers of coannihilating $N$ \cite{Klasen:2013jpa}.}
\end{figure}

If some of the singlet fermions have a mass slightly larger than that of $H^0$, coannihilations
with $N_i$ become relevant and give rise to an \emph{increase} in the relic density
(see Fig. \ref{fig:4}).
%
%
The relic density thus strongly depends on the mass difference between $H^0$ and the singlet
fermions.
The resulting indirect detection rate is large  and provides a  constraint on the parameter
space of the model (see Fig.\ \ref{fig:5}) \cite{Klasen:2013jpa}.
%
%
%
\begin{figure}
 \centering
 \includegraphics[width=1.0\columnwidth]{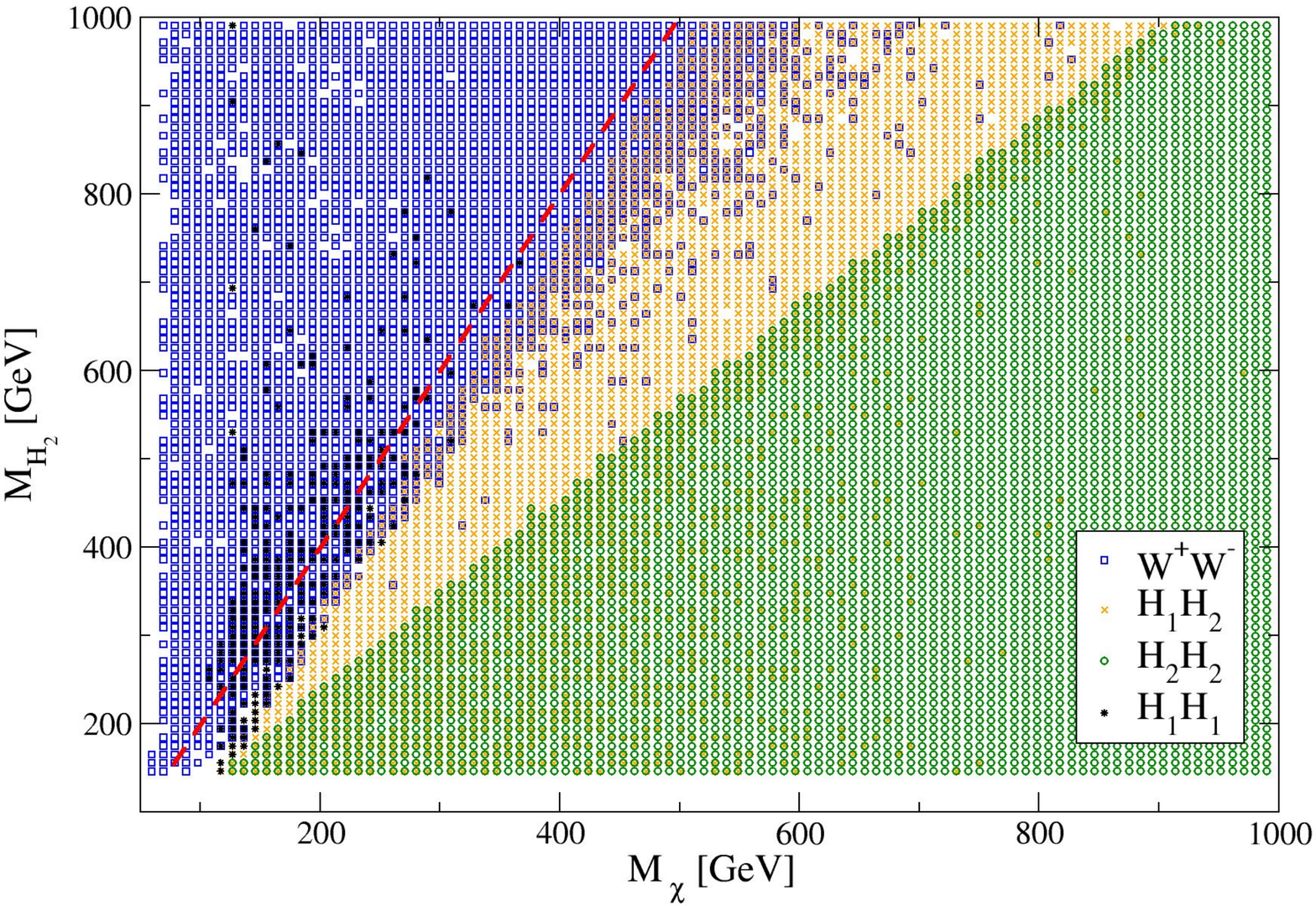}
 \caption{\label{fig:6}The region in the plane ($M_\chi$,$\mh$) that is compatible with the dark
 matter constraint. Different symbols are used to distinguish the dominant annihilation final
 states. The dashed (red) line shows the resonance condition: $2M_\chi=\mh$ \cite{Esch:2013rta}.}
\end{figure}
%

\subsection{The singlet fermionic model}
\label{}

\begin{figure}
 \centering
 \includegraphics[width=1.0\columnwidth]{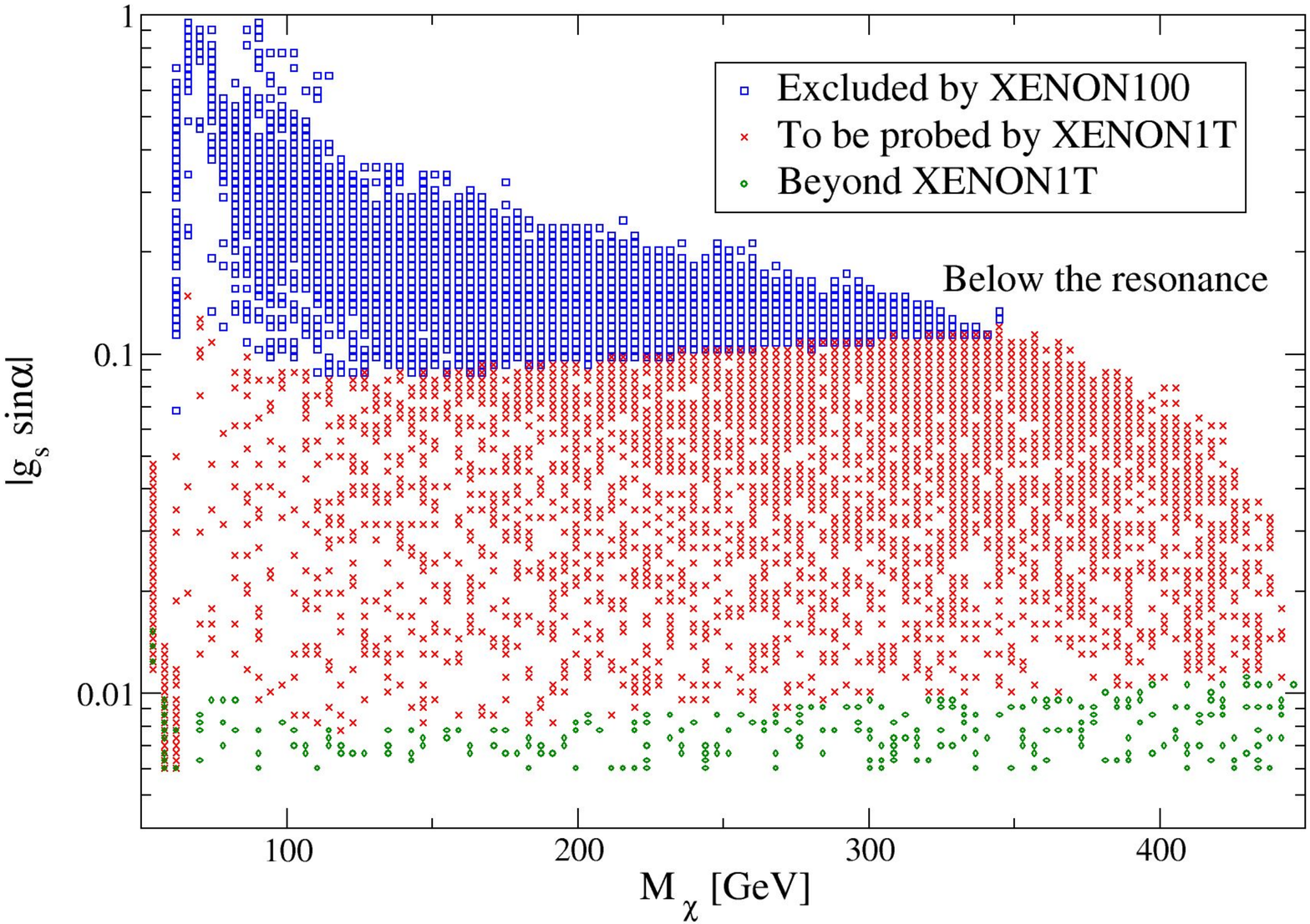}
 \caption{\label{fig:7}The reach of current and future direct detection experiments
 \cite{Aprile:2012zx} in the singlet fermion model below the resonance \cite{Esch:2013rta}.}
\end{figure}

In this model, the Standard Model is extended with a singlet fermion ($\chi$) and a singlet
scalar ($\phi$), which are odd and even under a $Z_2$, respectively. The fermion is therefore the
dark matter candidate and interacts via
\begin{equation}
 {\cal L}_\chi=g_s\phi\bar\chi\chi+ig_p\phi\bar\chi\gamma_5\chi.
\end{equation}
The new scalar mixes with the Higgs boson, giving rise to the mass eigenstates $H_1$ and $H_2$.
The dark matter annihilates mainly into four different channels: $W^+W^-$, $H_1H_1$, $H_1H_2$ and
$H_2H_2$ (see Fig.\ \ref{fig:6}). 
As one can see in Fig.\ \ref{fig:7}, some regions of the parameter space can already be
excluded by direct detection constraints. 
%
%


%
%

\section{Summary}
\label{}

Using three different examples for minimal extensions of the Standard Model, we have
illustrated in these proceedings the interplay of the Higgs boson discovery and dark
matter relic density constraints and their implications for direct and indirect searches
for dark matter.

\section*{Acknowledgments}

I thank S. Esch, D. Restrepo, J. Ruiz-Alvarez, C. Yaguna and O. Zapata for their collaboration.
Financial support by the Helmholtz Alliance for Astroparticle Physics and
the Deutsche Forschungsgemeinschaft under grant KL 1266/5-1 is gratefully
acknowledged.







\end{document}